\documentclass[pdflatex,sn-mathphys-num]{sn-jnl}

\usepackage{graphicx}%
\usepackage{multirow}%
\usepackage{amsmath,amssymb,amsfonts}%
\usepackage{amsthm}%
\usepackage{mathrsfs}%
\usepackage[title]{appendix}%
\usepackage{xcolor}%
\usepackage{textcomp}%
\usepackage{manyfoot}%
\usepackage{booktabs}%
\usepackage{algorithm}%
\usepackage{algorithmicx}%
\usepackage{algpseudocode}%
\usepackage{listings}%

\theoremstyle{thmstyleone}%
\theoremstyle{thmstyletwo}%

\theoremstyle{thmstylethree}%

\begin{document}

\title[Structural Balance in Real-World Social Networks: Incorporating Direction and Transitivity in Measuring Partial Balance]{Structural Balance in Real-World Social Networks: Incorporating Direction and Transitivity in Measuring Partial Balance}


\author*[1]{\fnm{Rezvaneh} \sur{Rezapour}}\email{shadi.rezapour@drexel.edu}\equalcont{These authors contributed equally to this work.}

\author[2]{\fnm{Ly} \sur{Dinh}}\email{lydinh@usf.edu}
\equalcont{These authors contributed equally to this work.}

\author[3]{\fnm{Lan} \sur{Jiang}}\email{lanj3@illinois.edu}

\author[4]{\fnm{Jana} \sur{Diesner}}\email{jana.diesner@tum.de}

\affil*[1]{\orgdiv{College of Computing and Informatics}, \orgname{Drexel University}, \orgaddress{\city{Philadelphia}, \state{PA}, \country{United States}}}

\affil[2]{\orgdiv{School of Information}, \orgname{University of South Florida}, \orgaddress{\city{Tampa}, \state{FL}, \country{United States}}}

\affil[3]{\orgdiv{School of Information Sciences}, \orgname{University of Illinois at Urbana-Champaign}, \orgaddress{ \city{Champaign}, \state{IL}, \country{United States}}}

\affil[4]{\orgdiv{School of Social Sciences and Technology}, \orgname{Technical University of Munich}, \orgaddress{ \city{Munich}, \country{Germany}}}


\abstract{Structural balance theory predicts that triads in networks gravitate towards stable configurations. The theory has been verified for undirected graphs. Since real-world networks are often directed, we introduce a novel method for considering both transitivity and sign consistency for evaluating partial balance in signed digraphs. We test our approach on graphs constructed by using different methods for identifying edge signs: natural language processing to infer signs from underlying text data, and self-reported survey data. Our results show that for various social contexts and edge sign detection methods, partial balance of these digraphs are moderately high, ranging from 61\% to 96\%. Our approach not only enhances the theoretical framework of structural balance but also provides practical insights into the stability of social networks, enabling a deeper understanding of interpersonal and group dynamics across different communication platforms.}




\maketitle

\section{Introduction}\label{sec1}

Social and communication networks in real world are composed of intricate and constantly changing interactions between individuals or groups. 
Network scholars have examined core principles that explain patterns of social interactions at various levels of analysis, i.e., node level \cite{borgatti2006graph}, dyadic level \cite{block2015reciprocity}, triadic level \cite{cartwright1956structural, johnsen1986structure}, subgroup level \cite{riley1986social}, and the graph level \cite{kilduff2003social}. 
In this study, we focus on a fundamental unit of analysis, namely triads, given that there are patterns of tie formations between three actors that cannot be explained at any other levels of analysis \cite{simmel1950stranger, flament1963applications, taylor1970balance}. 

Two fundamental theories that explain interactions at the triadic level are structural balance \cite{heider1946attitudes} and transitivity \cite{feld1982patterns}, which both model how stable relationships emerge in groups of three nodes. Theories of structural balance and transitivity have been applied to examine various social processes of opinion formation \cite{altafini2012dynamics}, cooperation in groups \cite{he2018evolution}, and maintenance of social status in organizations \cite{dong2015inferring}. There is also considerable empirical evidence in support of these theories \cite{chiang2019structural, diesner2015little, aref2020multilevel}, as well as evidence that falsifies the premises of the theories \cite{doreian2001pre, doreian2014testing,leskovec2010signed}. What these studies have in common is that they highlight the need to consider directionality in the evaluation of balance for networks where relationships are directed. Forcing a directed network to an undirected one for balance evaluation would remove additional configurations of balance between nodes \cite{song2015link,leskovec2010signed}, and exclude information about the direction(s) of influence between nodes \cite{csimcsek2020combined,uribe2020finding}. Thus, we take into account both the sign and direction of edges as properties to evaluate balance for ten real-world social networks, and empirically assess whether structural balance holds. Specifically, we assess partial balance \cite{aref2018measuring} for different types of signed directed relations, namely sentiment, morality, trust, friendship (preference), and alliance. Examining networks with various types of signed connections allows us to determine if equilibrium is maintained within and among different social relationships.

Our empirical assessment considers both transitivity and structural balance as two operating mechanisms for partial balance in signed and directed networks. This approach stems from the original formulation of balance \cite{heider1946attitudes, heider1958psychological} stating that transitivity is a crucial property, which explains how directed and signed ties are oriented in ways that are consistent with balance. We capture both transitive and balanced configurations by leveraging the triad census \cite{holland1971transitivity} to extract all transitive triples within a particular triad and calculate the overall balance with respect to the proportions of balanced triples within the triad. We also operationalize structural balance as partial balance, given that complete balance in real-world networks is rare \cite{aref2019balance,aref2018measuring}. To compute partial balance, we consider the relative frequencies of positive and negative triples within each triad, capturing the extent to which a network tends towards balance.
Altogether, we assess structural balance with transitivity and partial balance considered across the aforementioned signed relations and find that partial balance ratios vary, with an average balance ratio of 0.835. 

In comparison with the traditional method for determining balance, which utilizes an undirected approach, our findings indicate a more nuanced understanding of network dynamics. By incorporating directionality, we reveal hidden patterns of relationships and influences not captured in undirected models. Moreover, this directed approach helps in identifying asymmetric relationships, such as dominance or influence, which are critical in understanding social hierarchies and power dynamics. We also observe that different types of relationships exhibit distinct patterns of balance and transitivity, indicating the importance of context in network analysis. Our study contributes to a more comprehensive understanding of network theory by demonstrating the significance of considering both sign and direction in assessing the structural balance of real-world networks. This approach has potential applications in various domains, such as organizational behavior, social media analytics, and political science, where understanding the dynamics of complex relationships is crucial.

Our research marks a substantial advancement by empirically evaluating structural balance in a diverse range of real-world social relationships. We examine five principal types of social relationships, offering a comprehensive view of partial balance in various social networks. In addition, our findings reveal a critical insight: the coexistence and concurrent impact of sign consistency and transitivity in shaping the partial balance of networks. This finding highlights the intricate interplay between these mechanisms across different types of relationships, underscoring their pivotal role in network dynamics.

\section{Related work}
\begin{figure}[htb]
\centering
\centerline{\includegraphics[width=0.7\linewidth]{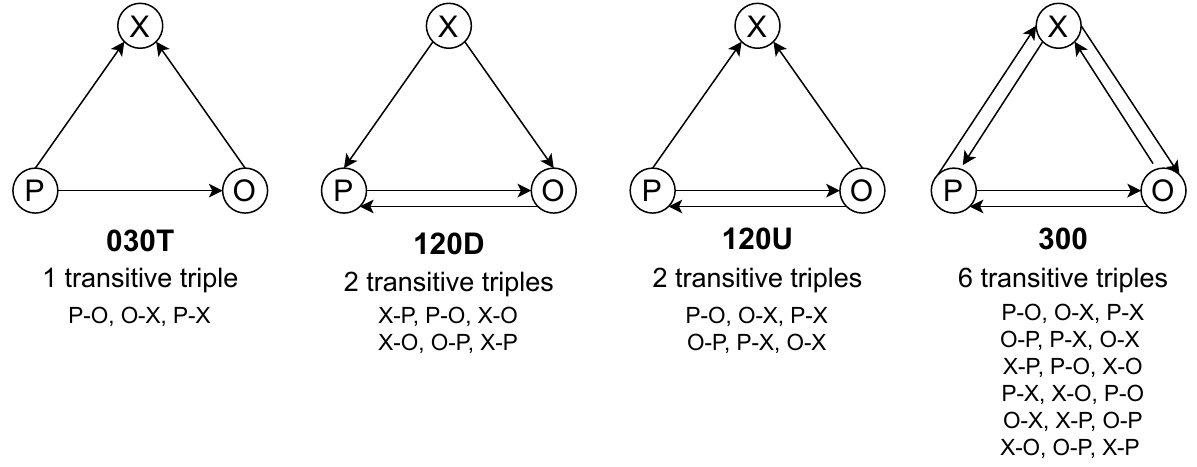}}
\caption{Triad census (include only triads which are transitive and balanced). Example of MAN census label - 030T: \textbf{M}utual=0, \textbf{A}symmetric=3, \textbf{N}ull=0, Letter Label: \textbf{T}=Transitive, \textbf{D}=Down, \textbf{U}=Up}
\label{fig:Fig1}
\end{figure}

The central tenet of social network analysis is to understand the structures of relations between sets of objects, such as individuals, groups, or organizations \cite{wasserman1994social}. Heider's structural balance theory \cite{heider1946attitudes} was the first formulation of how relationships form between three individuals, or between a pair of individuals, and their perceptions of or attitudes toward a common object. Heider initially examined $POX$ triads, where $P$ is a focal individual, $O$ is a second individual, and $X$ is either an individual or a common object. He asserted that the relationship between these three entities is `balanced' if $P$ liked $O$, $O$ liked $X$ and $P$ also liked $X$. On the other hand, `imbalance' would occur if $P$ liked $O$, $O$ liked $X$, but $P$ did not like $X$. Heider's primary claim was that balance is a state of equilibrium, and that individuals in networks strive to move towards and maintain that equilibrium or balance. Cartwright and Harary (1956) brought balance theory into the context of signed and directed networks, where relationships between pairs of nodes can be represented as either a ($+$) or ($-$). They further formalized that a triad within a network is balanced if the product of the signs of its edges is positive. Subsequently, Holland and Leinhart \cite{holland1971transitivity} introduced the transitivity model which asserted that transitivity is an important structural property of social relationships in general, and balance in particular. Specifically, Holland and Leinhart proposed eight possible types of triads for balance assessment, with transitivity considered (as shown in Figure \ref{fig:Fig1}). 

The relationship between transitivity and structural balance is explicitly stated in Heider's formulation as: ``three positive relations may be considered psychologically transitive.'' This means that ``$P$ tends to like $X$ if $P\mathcal{R}O$ and $O\mathcal{R}X$ are valid at the same time'' ($\mathcal{R}$ represents positive relation between two nodes) (\cite{heider1946attitudes}, p. 109). Thus, Heider suggested that beyond cognitive consistency as a primary theoretical mechanism, transitivity may also co-exist in a triad, though his formulation was at the level of an individual's perception. Holland and Leinhart \cite{holland1971transitivity} generalized this conjecture further to the context of a social arrangement with three individuals, represented as a triad of transitive relations (or a ``t-graph''). Feld and Elmore \cite{feld1982patterns}, followed by Doreian and Krackhardt \cite{doreian2001pre} found that differential popularity was another process, beyond cognitive consistency highlighted in Heider, that explained the presence of transitivity in balance. In fact, Doreian and Krackhardt \cite{doreian2001pre} found that transitivity and balance were both present, but rival processes in the Newcomb fraternity dataset \cite{newcomb1968interpersonal}. Hummon and Doreian \cite{hummon2003some} also found the complexity of the balance process to involve multiple operating mechanisms, namely transitivity, balance, and reciprocity. Davis, Holland, and Leinhardt \cite{davis1979davis} in their studies of positive relations in triads found that transitivity is a pervasive property in balanced (all positive) triads. Stix \cite{stix1974improved} also asserted that transitivity is an important property of balance, and that ``the transitivity of any structure corresponds to its sensitivity to imbalance'' (p. 447). Altogether, prior literature shows that the relationship between transitivity and balance is complex; it is still uncertain whether the two are complementary or rivaling processes. All we know is that they are distinctive but often co-existing in social relations \cite{stix1974improved, hummon2003some, holland1971transitivity}. 

Empirical studies on structural balance in real-world social networks offer mixed evidence regarding the theory. Several studies have found evidence that structural balance is present in email communication networks \cite{diesner2015little}, sentiment networks of individuals from the same (and different) nationalities \cite{chiang2019structural}, and ten real-world benchmark network datasets \cite{aref2020multilevel} based on several relationship types such as friendship, trust, alliance relations. In parallel, there are also a number of studies that did not find evidence for the theory \cite{doreian2019structural, doreian2001pre}. Doreian and Mrvar \cite{doreian2019structural} in their structural examination of signed and directed relations between countries from 1946-1999 found that the networks did not gravitate towards balance, and that balance was not achieved in the network (in contrary to Aref and Wilson \cite{aref2019balance}'s finding that this network achieves high degree of balance ($\gtrapprox$ 0.86). Rawlings and Friedkin \cite{rawlings2017structural} found that some statements of balance hold true (e.g., a friend of a friend is a friend), while others are violated (e.g., an enemy of an enemy is a friend) in the context of signed directed networks of sentiment among members in the same housing community. While evidence for structural balance has been mixed, these studies show that real-world relationships are complex, where properties such as signs and direction co-exist.  

In this study, we revisit the longstanding literature on transitivity and balance to calculate structural balance for networks of different signed relations. Specifically, we calculate balance, with transitivity considered, using Holland and Leinhart's triad census \cite{holland1971transitivity} and Estrada \cite{estrada2019rethinking} approach for analyzing triples. The triad census includes 16 classes of MAN (Mutual, Asymmetric, Null) triads, representing all possible combinations of directed ties between three nodes. Specifically, each configuration contains different combinations of edges, either mutual ($P$ likes $O$ and $O$ likes $P$), asymmetric ($P$ likes $O$ but $O$ does not like $P$), or null ($P$ and $O$ do not like each other). The triad census is especially useful for our calculation, as it contains four triad types that are driven by both transitivity and balance (as shown in Figure \ref{fig:Fig1}). Furthermore, the triad census includes all the possible triples (containing directed ties) embedded in each triad type, allowing us to conveniently extract transitive triples and exclude intransitive (cyclic) triples in our analyses. Another reason to exclude cyclic triples, besides the fact that they violate the transitivity rule, is that cycles contained limited information on the process of influence among relationships \cite{rank2010structural,veenstra2013network,evmenova2020analysis}. In the next section, we show that these four configurations are relevant for our operationalization of balanced triads with transitivity as a co-existing operating mechanism. 

\begin{figure}[htb]
\centering
\centerline{\includegraphics[width=0.5\linewidth]{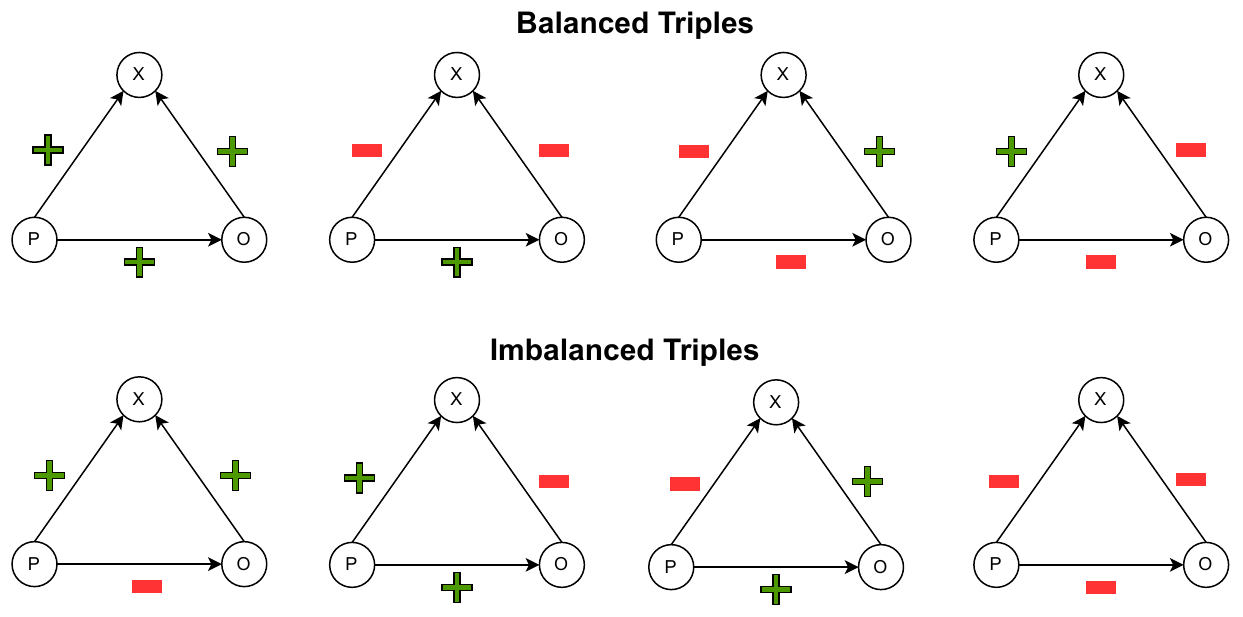}}
\caption{Balanced and imbalanced triples}
\label{fig:Fig2}
\end{figure}

\section{Balance evaluation for signed directed networks with transitivity considered} \label{RQ}
In this section, we present our approach to calculating partial balance in signed directed networks with respect to transitive triples (typologies of balanced and imbalanced transitive triples are shown in Figure \ref{fig:Fig2}).

\subsection{Problem definition} \label{theory}

\textbf{Definition 1}: 
Let $G$ be a signed digraph where $G = (D,\sigma)$. $D$ is a digraph underlying $G$, where ${D} = (V,E)$ and sign function $\sigma : E \rightarrow \{+,-\}$. A \textit{triad} $T$ in $G$ is a set of three distinct nodes with (at least) one directed edge between each two of them. 

A \textit{triple} $S$ in signed directed $T$ is a set of three directed edges that starts from a vertex $V$, follows the direction of edges, and does not return to the same vertex. In other words, $S$ is \textit{transitive} and \textit{non-cyclic}. Every \textit{triple} $S$ in a signed digraph $G$ must be \textit{transitive} in order to be considered for our balance analysis. 
Based on triad census \citep{holland1971transitivity}, we recognize four types of triads as transitive: (030T, 120D, 120U, and 300). These specific triad types are deemed transitive due to the presence of only transitive triples within them. Specifically, the 030T triad contains one transitive triple, while the 120D and 120U triads each contain two transitive triples, and the 300 triad contains a total of six transitive triples.


\textbf{Proposition:} We define $T$ a \textit{completely balanced} triad if and only if every transitive triple in $T$ is balanced. A transitive triple is \textit{positive} if it contains an even number of negative directed edges. Furthermore, we define $T$ a \textit{partially balanced} triad if it contains at least one negative triple. Finally, $T$ is \textit{completely imbalanced} if every transitive triple in $T$ is imbalanced (or negative thereof). 

After calculating balance $B_{T^{(i)}}^{j}$ for each triad $j$ of type $i$, we compute the weighted balance ratio for the set of all transitive triads of type $i$ ($B_{T^{(i)}}$). Finally, the \textit{overall balance ratio of $G$} ($B_{Avg(G)}$) is calculated by averaging the balance ratio of all triads across a network.  
A signed digraph $G = (D,\sigma)$ is balanced if all triads $T$ in $G$ are balanced. 


\textbf{Definition 2}: 
We compare our method for determining structural balance with the conventional approach, where networks are assumed to be undirected. 
We consider $G'$ an undirected signed network where $G' = (D',\sigma)$, and $D'$ is a graph underlying $G'$, where ${D'} = (V',E')$ and sign function $\sigma : E' \rightarrow \{+,-\}$. A \textit{triad} $T'$ in $G'$ is a set of three distinct nodes with (at least) one edge between each two of them. 
The sign of $T'$ is the product of the signs of its edges. A $T'$ is balanced if its sign is positive and unbalanced otherwise. After calculating balance for each $T'$, we compute the average balance ratio in $G'$. The \textit{overall balance ratio of $G'$} ($B_{Avg(G')}$) is calculated by averaging the balance ratio of all triads across a network.  
A signed graph $G' = (D',\sigma)$ is balanced if all triads $T'$ in $G'$ are balanced.

\section{Empirical Analysis}
\subsection{Data}
Several methods such as interviews, surveys, observations, and data mining are employed to gather social network data, enabling the construction and analysis of social connections and interactions among entities. Moreover, real-world communication, whether verbal or nonverbal, written or visual, encompasses a range of explicit and implicit relationships, including those characterized by like or dislike, trust or distrust.

For this study, we selected a wide range of datasets with different sizes (Table \ref{tab:Table2}) to better validate our proposed method for calculating balance in directed triads. Our selected datasets (ten datasets in total) cover a wide range of data collection techniques and portray multiple types of relationships in the networks; namely sentiment, morality, trust, friendship preference, and alliance. 
The following section provides more details about each dataset.

\vspace{0.5cm}
\noindent\textbf{Highland Tribes:}
The Highland tribes dataset represents an alliance structure among tribal groups using the Gahuku-Gama system of the Eastern Central Highlands of New Guinea. The dataset consists of two parts: (1) GAMAPOS representing alliance (``rova'') relations, and (2) GAMANEG for antagonistic (``hina'') relations \cite{read1954cultures,hage1983structural}.

\noindent\textbf{College House:}
The college A, B, and C datasets are constructed with respect to the relationships between three different groups of college girls at an Eastern college in the United States. Each group consists of approximately 20 members who lived together for at least four months. The girls in each house were asked to provide evaluations of other girls in the same house based on a range of behavioral characteristics. These data are then converted into matrices of choice and rejection for each house \cite{lemann1952group}.

\noindent\textbf{Bitcoin:}
The Bitcoin OTC and the Bitcoin Alpha datasets both represent trust between Bitcoin traders on two platforms, called Bitcoin OTC and Bitcoin Alpha. The transactions on these platforms are anonymous. However, to minimize fraudulent behavior and to maintain reputation, members of OTC and Alpha rate others in a scale of -10 (total distrust) to +10 (total trust) in steps of 1 \cite{kumar2016edge}.

\noindent\textbf{Enron:}
The Enron email data is a large-scale, temporal dataset which provides real-world organizational communication data of a global, U.S. based, former energy brokerage that went bankrupt in 2001 over a span of 3.5 years. The communication (email) dataset of 158 employees was released in 2002 by the FERC \cite{diesner2005communication, investigationpa02}. The original dataset went through various edits and modification over the years. In this study, we use the latest release of the dataset from 2015 ({\url{https://www.cs.cmu.edu/~./enron/}}). 

We disambiguated the email addresses of 558 Enron employees and converted them to actual names of the people in the Enron dataset \cite{diesner2005communication,diesner2015little}. 
Furthermore, we used natural language processing techniques to analyze emails and label each tie (email) based on their valence with respect to moral values \cite{graham2013moral,rezapour2019enhancing} and sentiment \cite{wiebe2005creating,jiang2020group}. More details on text analysis methods is provided in \S \ref{appendix_1}. 

\noindent\textbf{Avocado-IT:}
The Avocado Research Email Collection \cite{oard2015avocado} is provided by the Linguistic Data Consortium ({\url{https://catalog.ldc.upenn.edu/LDC2015T03}}) and consists of emails between 279 accounts of a defunct information technology company referred to as ``AvocadoIT'', a pseudonym assigned for anonymity. 
The dataset consists of calendars, attachments, contacts, reports, and emails. In this study, we only used the email communication for studying balance. 

For the Avocado dataset, we considered emails that were sent to or from corporate email addresses (emails ending to \textit{@avocadoit.com}). Similar to the Enron dataset, we further used text mining techniques to label each email based on their valence with respect to moral values \cite{graham2013moral,rezapour2019enhancing} and sentiment \cite{wiebe2005creating}. \S \ref{appendix_1} provides more details regarding the methods used for analyzing emails.

\subsection{Balance analysis based on edge signs and transitivity}

To analyze balance in triads, we follow the steps explained in \S \ref{theory}. Moreover, we used $NetworkX$, a $Python$ library, to extract instances of four transitive triads (030T, 120D, 120U, and 300) and analyze balance within each triad with respect to their triples. For networks with several components, we first selected the giant component and removed self-loops, pendants, and edges with neutral (zero) scores from the networks to make the analysis more efficient.


\begin{sidewaystable}
\begin{minipage}[]{15cm}
\centering
\begin{tabular}{|c|c|r|r|r|r|r|r|r|r|r|r|}
\hline
\multicolumn{1}{|c|}{\begin{tabular}[c]{@{}l@{}}Network Measures \end{tabular}} & 
\multicolumn{1}{l|}{\begin{tabular}[c]{@{}l@{}}Bitcoin-\\ OTC\end{tabular}} & \multicolumn{1}{l|}{\begin{tabular}[c]{@{}l@{}}Bitcoin-\\ Alpha\end{tabular}} & \multicolumn{1}{l|}{\begin{tabular}[c]{@{}l@{}}Highland\\ Tribes\end{tabular}} & \multicolumn{1}{l|}{\begin{tabular}[c]{@{}l@{}}College \\ House A\end{tabular}} & \multicolumn{1}{l|}{\begin{tabular}[c]{@{}l@{}}College\\ House B\end{tabular}} & \multicolumn{1}{l|}{\begin{tabular}[c]{@{}l@{}}College\\ House C\end{tabular}} & \multicolumn{2}{c|}{Enron} & \multicolumn{2}{c|}{Avocado} 
\\ \hline
Relational type 
& \multicolumn{1}{l|}{Trust} &  \multicolumn{1}{l|}{Trust} & \multicolumn{1}{l|}{Alliance} & \multicolumn{1}{l|}{Preference} & \multicolumn{1}{l|}{Preference} & \multicolumn{1}{l|}{Preference} & \multicolumn{1}{c|}{Morality} & \multicolumn{1}{c|}{Sentiment} & \multicolumn{1}{c|}{Morality} & \multicolumn{1}{c|}{Sentiment} 
\\ \hline
\# of nodes 
& 5881 & 3783 & 16 & 21 & 17 & 20 & 517 & 518 & 557 & 575 
\\ \hline
\# of edges 
& 35592 & 24186 & 116 & 94 & 83 & 81 & 7605 & 7510 & 22479 & 23910 
\\ \hline
\# of components 
& 4 & 5 & 1 & 1 & 1 & 1 & \multicolumn{2}{|c|}{1} & \multicolumn{2}{|c|}{19} 
\\ \hline
\multicolumn{11}{|c|}{Giant Component Statistics} \\ \hline 
\# of nodes 
& 5047 & 3381 & 16 & 21 & 17 & 20 & 494 & 491 & 452 & 402 
\\ \hline
\# of edges 
& 34759 & 23786 & 116 & 94 & 83 & 80 & 7520 & 7344 & 22953 & 23519 
\\ \hline
Transitivity 
& 0.05 & 0.07 & 0.53 & 0.39 & 0.40 & 0.27 & 0.21 & 0.2 & 0.5 & 0.5 
\\ \hline
Density 
& 0.001 & 0.002 & 0.48 & 0.22 & 0.31 & 0.21 & 0.031 & 0.03 & 0.11 & 0.14 
\\ \hline
Average Path Length 
& 3.44 & 3.49 & 1.54 & 2.21 & 1.95 & 2.34 & 2.53 & 2.56 & 1.7 & 1.6 
\\ \hline
Clustering Coefficient 
& 0.21 & 0.20 & 0.54 & 0.35 & 0.37 & 0.25 & 0.46 & 0.46 & 0.62 & 0.68 
\\ \hline
\end{tabular}
\caption{Descriptive network measures for all networks}
\label{tab:Table2}
\end{minipage}
\vspace*{1 cm}

\begin{minipage}[]{15cm}
\centering
\begin{tabular}{|l|rrrrrrrrrr|}
\hline
 & \multicolumn{10}{c|}{Balance ratio (and counts of triads) for each triad type} \\ \hline
Type 
& \multicolumn{1}{c|}{\begin{tabular}[c]{@{}c@{}}Bitcoin-\\ OTC\end{tabular}} & \multicolumn{1}{c|}{\begin{tabular}[c]{@{}c@{}}Bitcoin-\\ Alpha\end{tabular}} & \multicolumn{1}{c|}{\begin{tabular}[c]{@{}c@{}}Highland\\ Tribes\end{tabular}} & \multicolumn{1}{c|}{\begin{tabular}[c]{@{}c@{}}College\\ House A\end{tabular}} & \multicolumn{1}{c|}{\begin{tabular}[c]{@{}c@{}}College\\ House B\end{tabular}} & \multicolumn{1}{c|}{\begin{tabular}[c]{@{}c@{}}College \\ House C\end{tabular}} & \multicolumn{1}{c|}{\begin{tabular}[c]{@{}c@{}}Enron-\\ Morality\end{tabular}} & \multicolumn{1}{c|}{\begin{tabular}[c]{@{}c@{}}Enron-\\ Sentiment\end{tabular}} & \multicolumn{1}{c|}{\begin{tabular}[c]{@{}c@{}}Avocado-\\ Morality\end{tabular}} & \multicolumn{1}{c|}{\begin{tabular}[c]{@{}c@{}}Avocado-\\ Sentiment\end{tabular}} 
\\ \hline
030T 
& \multicolumn{1}{r|}{\begin{tabular}[c]{@{}r@{}}0.91\\ (3,706)\end{tabular}} & \multicolumn{1}{r|}{\begin{tabular}[c]{@{}r@{}}0.82\\ (974)\end{tabular}} & \multicolumn{1}{r|}{\begin{tabular}[c]{@{}r@{}}0.00\\ (0)\end{tabular}} & \multicolumn{1}{r|}{\begin{tabular}[c]{@{}r@{}}0.67\\ (27)\end{tabular}} & \multicolumn{1}{r|}{\begin{tabular}[c]{@{}r@{}}0.43\\ (21)\end{tabular}} & \multicolumn{1}{r|}{\begin{tabular}[c]{@{}r@{}}0.82\\  (17)\end{tabular}} & \multicolumn{1}{r|}{\begin{tabular}[c]{@{}r@{}}0.91\\  (4,514)\end{tabular}} & \multicolumn{1}{r|}{\begin{tabular}[c]{@{}r@{}}0.67\\  (4,238)\end{tabular}} & \multicolumn{1}{r|}{\begin{tabular}[c]{@{}r@{}}0.81\\  (8,787)\end{tabular}} & \multicolumn{1}{r|}{\begin{tabular}[c]{@{}r@{}}0.76\\  (8,577)\end{tabular}} 
\\ \hline
120D 
& \multicolumn{1}{r|}{\begin{tabular}[c]{@{}r@{}}0.85\\ (2,096)\end{tabular}} & \multicolumn{1}{r|}{\begin{tabular}[c]{@{}r@{}}0.82\\  (1,142)\end{tabular}} & \multicolumn{1}{r|}{\begin{tabular}[c]{@{}r@{}}0.00\\ (0)\end{tabular}} & \multicolumn{1}{r|}{\begin{tabular}[c]{@{}r@{}}0.85\\ (13)\end{tabular}} & \multicolumn{1}{r|}{\begin{tabular}[c]{@{}r@{}}0.33\\ (6)\end{tabular}} & \multicolumn{1}{r|}{\begin{tabular}[c]{@{}r@{}}1.00\\  (8)\end{tabular}} & \multicolumn{1}{r|}{\begin{tabular}[c]{@{}r@{}}0.92\\  (2,390)\end{tabular}} & \multicolumn{1}{r|}{\begin{tabular}[c]{@{}r@{}}0.68\\  (2,384)\end{tabular}} & \multicolumn{1}{r|}{\begin{tabular}[c]{@{}r@{}}0.86\\  (14,111)\end{tabular}} & \multicolumn{1}{r|}{\begin{tabular}[c]{@{}r@{}}0.81\\  (14,276)\end{tabular}} 
\\ \hline
120U 
& \multicolumn{1}{r|}{\begin{tabular}[c]{@{}r@{}}0.83\\ (2,910)\end{tabular}} & \multicolumn{1}{r|}{\begin{tabular}[c]{@{}r@{}}0.87\\  (1,780)\end{tabular}} & \multicolumn{1}{r|}{\begin{tabular}[c]{@{}r@{}}0.00\\  (0)\end{tabular}} & \multicolumn{1}{r|}{\begin{tabular}[c]{@{}r@{}}1.00\\  (13)\end{tabular}} & \multicolumn{1}{r|}{\begin{tabular}[c]{@{}r@{}}0.80\\  (15)\end{tabular}} & \multicolumn{1}{r|}{\begin{tabular}[c]{@{}r@{}}1.00\\  (3)\end{tabular}} & \multicolumn{1}{r|}{\begin{tabular}[c]{@{}r@{}}0.92\\  (3,615)\end{tabular}} & \multicolumn{1}{r|}{\begin{tabular}[c]{@{}r@{}}0.64\\  (3,513)\end{tabular}} & \multicolumn{1}{r|}{\begin{tabular}[c]{@{}r@{}}0.87\\  (26,165)\end{tabular}} & \multicolumn{1}{r|}{\begin{tabular}[c]{@{}r@{}}0.83\\  (28,615)\end{tabular}} 
\\ \hline
300 
& \multicolumn{1}{r|}{\begin{tabular}[c]{@{}r@{}}0.93\\ (13,752)\end{tabular}} & \multicolumn{1}{r|}{\begin{tabular}[c]{@{}r@{}}0.92\\  (9,894)\end{tabular}} & \multicolumn{1}{r|}{\begin{tabular}[c]{@{}r@{}}0.87\\  (68)\end{tabular}} & \multicolumn{1}{r|}{\begin{tabular}[c]{@{}r@{}}1.00\\  (4)\end{tabular}} & \multicolumn{1}{r|}{\begin{tabular}[c]{@{}r@{}}0.88\\  (4)\end{tabular}} & \multicolumn{1}{r|}{\begin{tabular}[c]{@{}r@{}}1.00\\ (1)\end{tabular}} & \multicolumn{1}{r|}{\begin{tabular}[c]{@{}r@{}}0.94\\ (3,056)\end{tabular}} & \multicolumn{1}{r|}{\begin{tabular}[c]{@{}r@{}}0.70\\  (3,056)\end{tabular}} & \multicolumn{1}{r|}{\begin{tabular}[c]{@{}r@{}}0.93\\  (124,371)\end{tabular}} & \multicolumn{1}{r|}{\begin{tabular}[c]{@{}r@{}}0.90\\  (144,865)\end{tabular}} 
\\ \hline
Average Balance (Directed) 
& \multicolumn{1}{r|}{\begin{tabular}[r]{@{}r@{}}0.88\\ (22,464)\end{tabular}} & \multicolumn{1}{r|}{\begin{tabular}[c]{@{}r@{}}0.86\\  (13,790)\end{tabular}} & \multicolumn{1}{r|}{\begin{tabular}[c]{@{}r@{}}0.87\\  (68)\end{tabular}} & \multicolumn{1}{r|}{\begin{tabular}[c]{@{}r@{}}0.88\\  (57)\end{tabular}} & \multicolumn{1}{r|}{\begin{tabular}[c]{@{}r@{}}0.61\\  (46)\end{tabular}} & \multicolumn{1}{r|}{\begin{tabular}[c]{@{}r@{}}0.96\\ (29)\end{tabular}} & \multicolumn{1}{r|}{\begin{tabular}[c]{@{}r@{}}0.92\\ (13,575)\end{tabular}} & \multicolumn{1}{r|}{\begin{tabular}[c]{@{}r@{}}0.68\\  (13,191)\end{tabular}} & \multicolumn{1}{r|}{\begin{tabular}[c]{@{}r@{}}0.87\\  (173,434)\end{tabular}} & \multicolumn{1}{r|}{\begin{tabular}[c]{@{}r@{}}0.82\\  (196,333)\end{tabular}} 
\\ \hline
\end{tabular}
\caption{Balance counts for different triad types across all networks}
\label{tab:Table3}
\end{minipage}
\end{sidewaystable}

\section{Results}
\subsection{Descriptive network measures}
Table \ref{tab:Table2} shows the structural characteristics of the ten networks with regard to the network size, transitivity, density, average path length, and clustering coefficient. We focus on the giant component of each network, meaning that pendants and isolates are excluded from analysis. The transitivity score is the highest for the Highland tribes network (0.53) and lowest for the Bitcoin-OTC network (0.05). High transitivity means the network has a sizable proportion of triads for balance analysis, whereas low transitivity suggests the network contains other configurations beyond the triad (e.g., cycles of length	$>$ 3). Density captures the occurrences of dyads in the network, and we find a close connection between density and transitivity scores. Highland tribes has the highest density (0.48).
This reveals the connection between the number of dyads and triads present in the network and how they allude to the interconnectedness between nodes. Measures of average path length and clustering coefficient reveal the topological characteristic of a network with respect to the presence of subgroups and communities. Most networks are found to have low average path length and mid-range clustering coefficient, meaning that they are exhibiting some characteristic of a small-world topology (low average path length, high clustering coefficient), though this needs to be statistically evaluated. Highland tribes network, Enron-morality, Avocado-morality, and Avocado-sentiment are three networks that most clearly exhibit the small-world effect.

\subsection{Partial balance analysis}
The average partial balance ratio across all networks (shown in Table \ref{tab:Table3}) is 0.835 (min = 0.61, max = 0.96, s.d = 0.10). The network with the highest partial balance is College House C (0.96), and the network with the lowest partial balance is College House B (0.61). We observe that high partial ratios persist across the different network sizes, with the smallest network having a balance ratio of 0.87 (Highland tribes), and the largest network having a balance ratio of 0.88 (Bitcoin OTC). 
We also examine the proportions of triads with respect to their balance ratios, as shown in Table \ref{tab:Table3}. For five of the networks (Bitcoin-OTC, Bitcoin-Alpha, Highland tribes, Avocado-morality, Avocado-sentiment), triad 300 is the most frequently occurring, and often most balanced when compared to the other triad types. The prevalence of balanced triads 300s shows that balance is often present in situations where individuals initiate and reciprocate the connection, specifically in the cases of trust (Bitcoin networks) and email communication (Avocado networks). For the other networks (College Houses A, B, C, Enron-morality, Enron-sentiment), triad type 030T is the most frequently occurring. The frequent presence of 030T in these networks shows that some relationships are bounded by a ``local hierarchy'' - in the context of Enron Sentiment, individual $P$ sends an email with a positive valance to $O$, $O$ replies to $X$ negatively, thus $P$ replies to $X$ negatively based on the prior interaction between $O$ and $X$. The prevalence of 030T triads in these particular networks also shows that reciprocity is lower for certain types of relationships, specifically 
friendship preferences, and email communication). On the other hand, trust relations as observed for the Bitcoin networks have a higher frequency of reciprocation within triads.  

\begin{table}
\centering
\resizebox{\textwidth}{!}{%
\begin{tabular}{|c|c|c|c|c||c|c|c|c|} \hline 
 & \multicolumn{4}{|c||}{Directed}& \multicolumn{4}{|c|}{Undirected}\\
\hline
\multicolumn{1}{|c|}{\textbf{}} & \multicolumn{1}{|c|}{+ + +} & \multicolumn{1}{|c|}{+ - -} & \multicolumn{1}{|c|}{+ + -} & \multicolumn{1}{|c||}{- - -} &  + + +& + - -& + + -&- - -\\ \hline
\textbf{Bitcoin- OTC} & 0.85 & 0.07 & 0.07 & 0.004  & 0.71& 0.16& 0.12&0.01\\ \hline
\textbf{Bitcoin-Alpha} & 0.87 & 0.04 & 0.08 & 0.003  & 0.78& 0.08& 0.14&0.01\\ \hline
\textbf{Highland Tribes} & 0.28 & 0.59 & 0.03 & 0.10  & 0.28& 0.59& 0.03&0.10\\ \hline
\textbf{College House A} & 0.42 & 0.45 & 0.05 & 0.08  & 0.23
& 0.48
& 0.14
&0.15
\\ \hline
\textbf{College House B} & 0.20 & 0.48 & 0.15 & 0.17  & 0.08
& 0.44
& 0.23
&0.26
\\ \hline
\textbf{College House C} & 0.33 & 0.60 & 0.02 & 0.05  & 0.19& 0.69& 0.04&0.08\\ \hline
\textbf{Enron Morality} & 0.92 & 0.01 & 0.07 & 0.0003  & 0.93& 0.00& 0.06&0.00\\ \hline
\textbf{Enron Sentiment} & 0.58 & 0.11 & 0.30 & 0.01  & 0.39& 0.39& 0.21&0.01\\ \hline
\textbf{Avocado Morality} & 0.92 & 0.01 & 0.07 & 0.0002  & 0.89
& 0.01
& 0.10
&0.00
\\ \hline
\textbf{Avocado Sentiment} & 0.89 & 0.01 & 0.10 & 0.004  & 0.85& 0.01& 0.13&0.00\\ \hline
\end{tabular}
}
\caption{Types of signed triples for all networks}
\label{tab:Table4}
\end{table}

\subsection{Sign analysis of triples}
All networks in our dataset contain higher proportions of positive than negative edges in both directed and undirected networks (results shown in Table \ref{tab:Table4}). For the majority of directed networks, all-positive triples are most prevalent, indicating the dominance of strong balance within the network. This result is consistent with Leskovec, Huttenlocher, and Kleinberg \cite{leskovec2010signed}'s finding that real-world social networks contain 70\% to 87\% of all-positive triples. The prevalence of strongly-balanced triples ($+++$) also show evidence in support of prior work by Doreian and Krackhardt \cite{doreian2001pre} and Davis \cite{davis1979davis}, who all found transitivity to be a pre-condition for balance when both $P$ $\rightarrow$ $O$ and $O$ $\rightarrow$ $X$ are positive. Our findings also show two possible outcomes for the pre-transitive condition; there is a higher likelihood that $P$ $\rightarrow$ $X$ will be positive, but there is also a possibility for $P$ $\rightarrow$ $O$ to be negative. For the remaining four networks (Highland Tribes, College House A, B, and C), there is a higher frequency of ($+--$) triples, showing empirical evidence of signed transitivity \cite{doreian2001pre} in several real-world networks. These four networks are all noticeably smaller than other networks, which may suggest that the prominence of ($+--$) triples may be specific to these social contexts, as opposed to being generalizable to other networks with the same relational type.  

For the imbalanced triples, we find that in six out of ten networks, ($++-$) triples are more prevalent than ($---$) triples. This finding is in contrast with \cite{szell2010multirelational}'s results, which found that ($++-$) triples are the least prominent in real-world networks, followed by ($---$) triples. Leskovec et al. \cite{leskovec2010signed} also empirically observed the rarity of ($++-$) triples in real networks. However, in the case of directed networks where edges are defined in terms of social status theory, ($++-$) triples are over-represented beyond what is expected by chance. Easley and Kleinberg \cite{easley2010chapter} also suggested that these two types of tensions are unique, and suggest that ($++-$) triples are more likely to be resolved over time. In the context of College House friendship preference, for example, where $A$ likes $B$, $B$ likes $C$, but $A$ does not like $C$. Person $A$ may be more likely to reconcile this tension by liking $C$. However, if there are all negative relations ($---$) between $A$, $B$, and $C$, Davis \cite{davis1967clustering} argues that it is less likely for two enemies to like each other. Instead, the three enemies are likely to just stay enemies, and not be involved in the triadic relationship at all. In the remaining four networks (Highland tribes, College Houses A, B, and C), we find slightly higher occurrences of one-negative than all-negative triads. This finding signals the difficulty for the triads to reconcile the tensions, and thus may be less likely to become more balanced over time. This is a particularly interesting finding for the case of Highland Tribes, showing that the alliance structure is rigid and less likely to change over time. In the context of three College House friendship networks, friendship preference is not likely to change, despite the fact that the change could reduce tension for the network as a whole.

\begin{figure}[ht]
\centering
\centerline{\includegraphics[width=\textwidth]{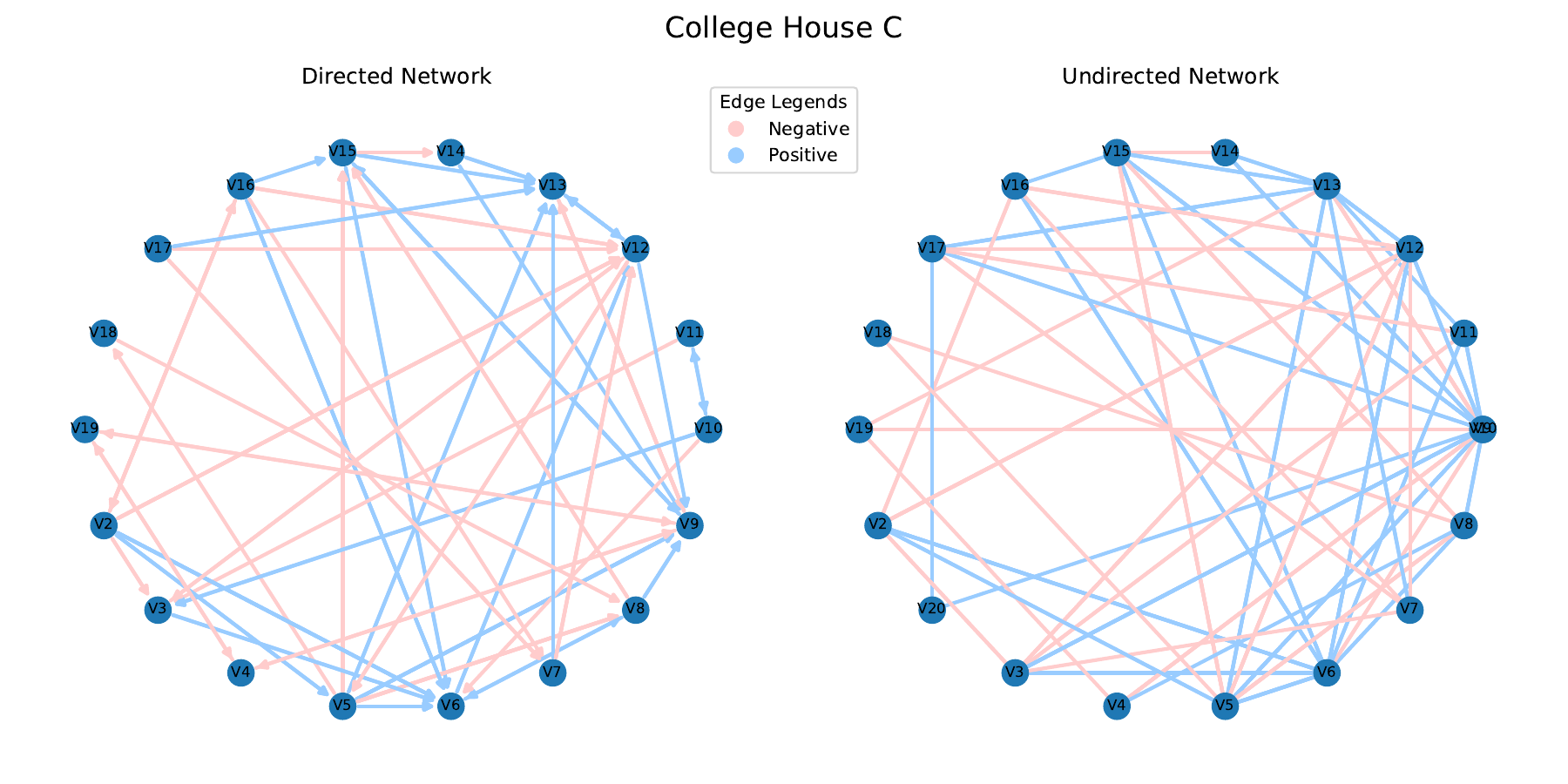}}
\caption{Case study: College House C network as a directed network (left) and undirected network (right)}
\label{fig:casestudy_CollegeC}
\end{figure}
Szell et al.'s \cite{szell2010multirelational} study brings out an important distinction between strong and weak structural balance. Weak structural balance \cite{davis1967clustering} posits that the ($++-$) triple is the only configuration that is (weakly) imbalanced, while the ($---$) is considered (weakly) balanced because empirical observations have found that ``an enemy of one's enemy can also be an enemy.'' Additionally, the all-negative triple is more likely to become balanced, than the one-negative triple, because it is more likely that two enemies will become friends because they have a mutual enemy (``an enemy of my enemy is my friend''). However, as alluded to in Easley and Kleinberg \cite{easley2010chapter}, the presence of the one-negative triple ($++-$) can also signal the possibility of resolving tension among a friend of a friend. For instance, in the Enron email network, we observe that $P$ may send a positive email to $X$ in the future, driven by the positive connection that $P$ has with $O$, and knowing that $O$ sent a positive email to $X$. Differing findings from the literature suggest that the sign configurations of triples are complex, especially when directed edges are considered \cite{leskovec2010signed}. In our analysis of these signed and directed triples, we find that weak balance is not prominent in relationships that are transitive. In other words, ``an enemy of an enemy is an enemy'' is not as likely to happen as a ``a friend of a friend is an enemy'' connection. It would be worthwhile to examine the temporal changes of these two triple types over time, and whether it is more likely that an enemy would become a friend in the ($---$) case, or that (1) an enemy would become a friend, or (2) a friend becomes an enemy in the ($++-$) triple. From this perspective, there are two possibilities for the ($++-$) to obtain balance, as opposed to ($---$) triple, which can only be resolved if an edge becomes positive.

Ultimately, the differences in these results show that (1) the composition of triples differs across different real-world contexts and relations, and (2) across directed and undirected realizations of the networks.

\begin{figure}[H]
\centering
\centerline{\includegraphics[width=\textwidth]{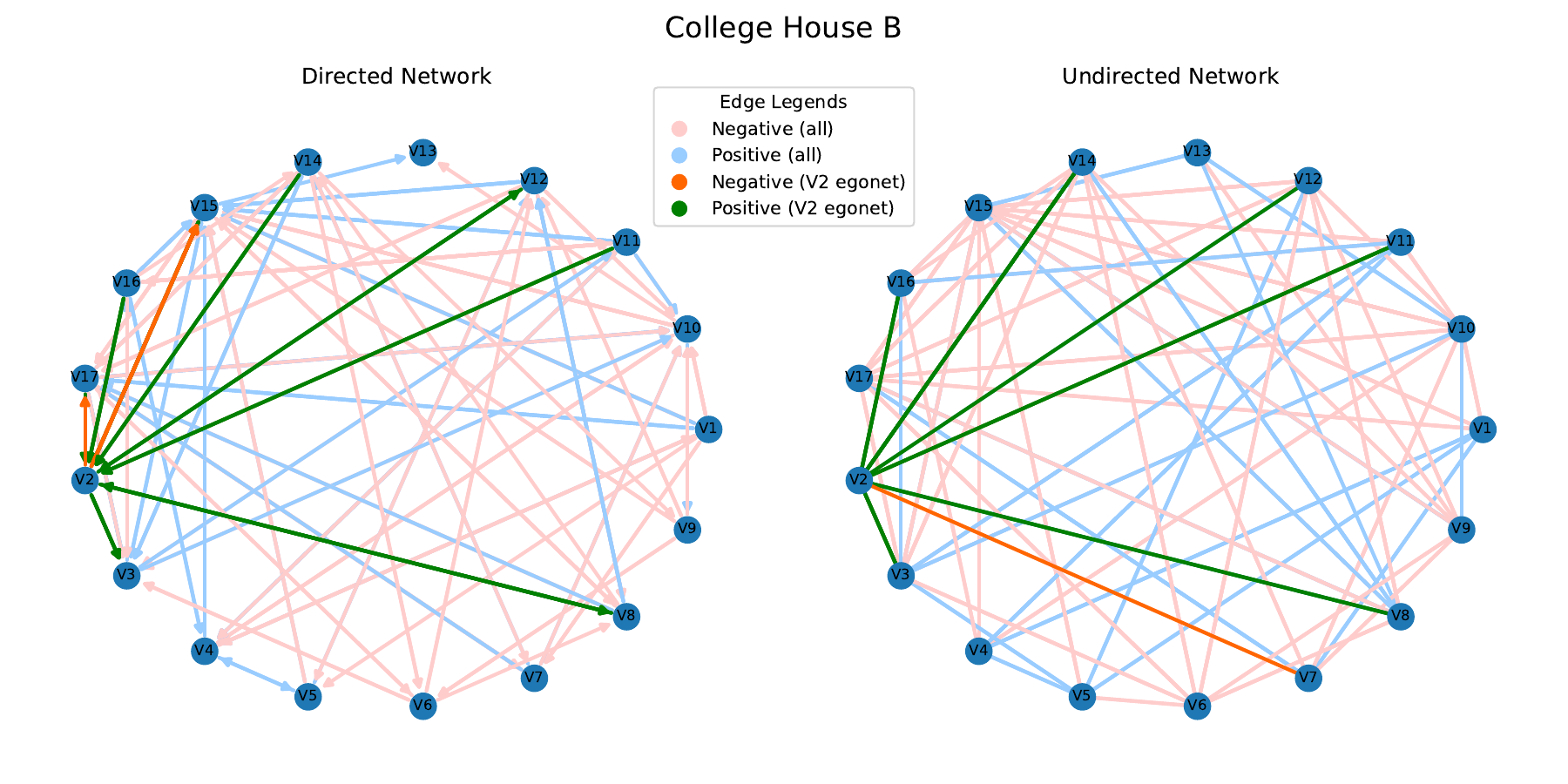}}
\caption{Case study: College House B network as a directed network (left) and undirected network (right). Edge Legends: \textit{all} indicates the whole network, and \textit{V2 egonet} indicates the egonet of node V2.}
\label{fig:casestudy_CollegeB}
\end{figure}

\subsection{Utility of directed networks}
The overall trend observed across all networks in our study is that the undirected realizations inflate the count of edges, triples, and consequently, the triads available for analysis. This inflation occurs because undirected networks do not require transitivity, meaning that not all connections must follow a certain pattern of relationships, making more configurations available for analysis. For instance, in College House A and College House C networks, the undirected realizations included a notable number of extra triples which subsequently influenced the balance score. In particular, the undirected realization of College House A contains 48 extra triples than the directed version, in which 24 are balanced in sign products, and 24 are imbalanced. Hence, the resulting balance score is influenced and dropped to 0.72 as opposed to 0.88 in the directed network. The College House C network (shown in Figure \ref{fig:casestudy_CollegeC} exhibits similar patterns as the College House A network in terms of the inflated number of triples present in the undirected network. There are 19 additional triples not presented in the directed network, 16 of which are balanced, and 3 are imbalanced. For instance, there is a triad that exists between nodes V4, V9, and V10 in the undirected network but not the directed network because this particular triad violates the transitivity assumption. In particular, this triad is a cycle, whereby V4 $\rightarrow$ V9 (+ sign), V9 $\rightarrow$ V10 (-), and V10 $\rightarrow$ V4 (+). Similar situation occurred with triad V3, V6, V12 where the triad contains one cycle (V3 $\rightarrow$ V6 (-), V6 $\rightarrow$ V12 (-), V12 $\rightarrow$ V3 (+)). This particular triad, however, also contains one triad that satisfies both transitivity and balance conditions (V3 $\rightarrow$ V6 (-), V6 $\rightarrow$ V12 (-), V3 $\rightarrow$ V12 (+)), but the presence of a violating cycle influences the overall transitivity of the triad. 

\begin{figure}[htb]
\centering
\centerline{\includegraphics[width=0.7\linewidth]{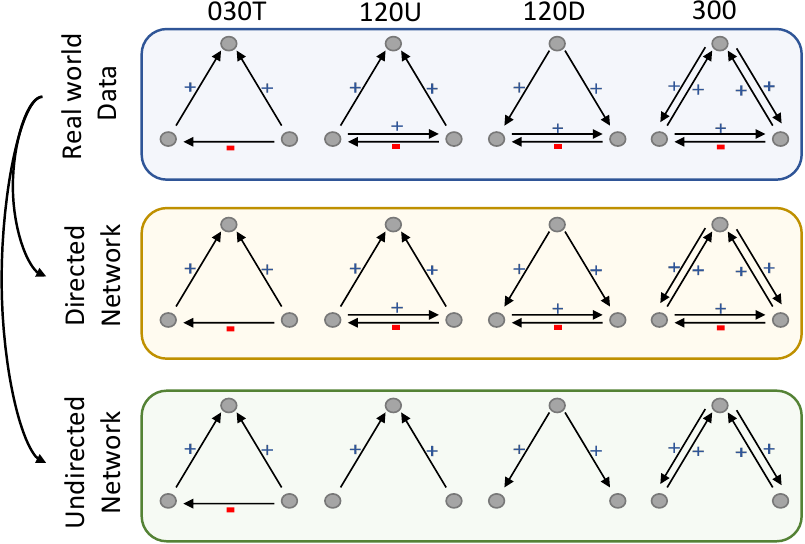}}
\caption{Examining differences in network analysis outcomes with and without considering directionality in real-world data}
\label{fig:Fig4}
\end{figure}

Undirected realizations also remove edges where there is a discrepancy in the signs of the incoming and outgoing edges between two nodes, as shown in Figure \ref{fig:Fig4}). In the College House B network, we observe the loss of three particular edges that in turn influence the final balance score of the undirected realization. As shown in Figure \ref{fig:casestudy_CollegeB}, two edges V2 $\rightarrow$ V15 (-), and V2 $\rightarrow$ V17 (-) from the \textit{V2 egonet} are present in the directed network but not the undirected network, as they were canceled out due to the mismatch in edge signs. The directed realizations hence manage to preserve at least an edge between V2 and V15, for instance, because they are involved in multiple triadic configurations that satisfy the transitivity and balance assumptions in the directed network (e.g., V15, V11, V2 as 120D triad; V3, V15, V2 as 120U triad). There is an exception to V2 $\rightarrow$ V7 edge being present in the undirected network but not the directed network. This is because, in the directed network, this edge is not part of any triads that meet the transitivity requirement. On the other hand, the undirected network keeps this edge, but this is problematic because there is only an edge between V2 $\rightarrow$ V7, but not V7 $\rightarrow$ V2. 

The Highland Tribes network stands out as the only instance where the directed and undirected networks share identical triad composition and overall balance scores. This is because the network is completely connected, hence there is a maximal number of triads that can exist in both network realizations. Such a network structure is, however, rare in real-world networks. Our case study of College Houses A, B, and C demonstrates that even the slightest discrepancy in the edge configurations and their signs when the network is switched from directed to undirected network can significantly impact the network's structure and overall balance score.

\begin{table}
\centering
\resizebox{\textwidth}{!}{%
\begin{tabular}{|l|r|r|r|r|r|r|r|r|r|} 
\hline
 & \multicolumn{3}{c|}{Directed\_Partial} & \multicolumn{3}{c|}{Directed\_Non-Partial} & \multicolumn{3}{c|}{Undirected} \\ 
\hline
 & \multicolumn{1}{l|}{BR} & \multicolumn{1}{l|}{\# BT} & \multicolumn{1}{l|}{\# IT} & \multicolumn{1}{l|}{BR} & \multicolumn{1}{l|}{\# BT} & \multicolumn{1}{l|}{\# IT} & \multicolumn{1}{l|}{BR} & \multicolumn{1}{l|}{\# BT} & \multicolumn{1}{l|}{\# IT} \\ 
\hline
College House A & 0.88 & 46 & 11 & 0.8 & 46 & 11 & 0.74 & 29 & 10 \\ 
\hline
College House B & 0.61 & 29 & 17 & 0.52 & 24 & 22 & 0.44 & 14 & 18 \\ 
\hline
College House C & 0.96 & 26 & 3 & 0.89 & 26 & 3 & 0.92 & 25 & 2 \\ 
\hline
Bitcoin-Alpha & 0.86 & 13069 & 721 & 0.84 & 11649 & 2141 & 0.84 & 2930 & 546 \\ 
\hline
Bitcoin-OTC & 0.88 & 21192 & 1272 & 0.86 & 19447 & 2969 & 0.91 & 3814 & 360 \\ 
\hline
Highland Tribes\_sym & 0.87 & 59 & 9 & 0.87 & 59 & 9 & 0.92 & 24 & 2 \\ 

\hline
Enron sentiment & 0.68 & 10444 & 2747 & 0.53 & 6045 & 7165 & 0.69 & 2104 & 949 \\ 
\hline
Avocado\_sentiment & 0.82 & 185832 & 10501 & 0.46 & 105222 & 91505 & 0.85 & 8846 & 1495 \\ 
\hline
Avocado\_morality & 0.87 & 165639 & 7795 & 0.63 & 117515 & 67781 & 0.84 & 8060 & 1544 \\ 
\hline
Enron morality & 0.92 & 12856 & 719 & 0.79 & 10908 & 2813 & 0.93 & 2831 & 226 \\
\hline
\end{tabular}}
\caption{Comparing Partial and Non-Partial Balance. BR: Balance ratio, BT: balance triangles, IT: imbalance triangles}
\end{table}

\section{Discussion and Conclusion}
\label{S:2}
In this paper, we employ an empirical framework for calculating balance in signed, transitive digraphs, which was essential to appropriately model and study balance in real-world communication networks and other networks where links might be asymmetric. We applied this framework to ten social networks with five different relational types (trust, alliance, friendship preference, morality, and sentiment). 
Our rationale for testing our approach on different networks was to determine whether mechanisms of structural balance and transitivity hold true across diverse social contexts. Moreover, prior research has mainly examined structural balance in signed and undirected graphs. Our study provided an actionable solution to measure structural balance in signed digraphs, using principles of transitivity to evaluate the directionality between edges. 

Overall, our findings showed that the amount to which a network was balanced was strongly impacted by choices of measuring social relations. When the direction of edges was taken into account, along with sign consistency, we expected that the overall balance ratio may be different than findings where only sign consistency was considered \cite{diesner2015little,leskovec2010signed}. Choices of relational type may also affect the overall partial balance. Our findings showed that each relational type captured a different characteristic of a network, as reflected in the different balance ratios across dimensions of social trust, friendship preference, alliance relations, morality, and sentiment of email (and online) communications. While partial balance ratios for all relational types were about 0.82 and above (balance higher than imbalance), we found the compositions of triples that make up balanced triads are different depending on the type of relationship. 

The patterns of partial balance that we discovered across the ten real-world networks offer implications for existing communication and social networks literature. First, we found that partial balance with transitivity considered provides nuances into the co-existence of hierarchy, influence, and reciprocity as mechanisms that influence balance. For instance, if we consider an undirected and balanced triad ($ABC$) where we know $A$-$B$ is positive, $B$-$C$ is positive, and $C$-$A$ is positive, the lack of edge direction does not explain the potential source(s) of balance. Whereas if we are presented with additional information, such as $A$ $\rightarrow$ $B$ is positive, $B$ $\rightarrow$ $C$ is positive, $A$ $\rightarrow$ $C$ is positive, we observe that $A$ likely has a positive edge towards $C$ because $B$ has a positive edge for $C$, hence the process of social influence is captured. Secondly, our partial balance and sign analyses reveal that triad type 300 is most prominent in all empirical networks, and specifically this triad 300 is often composed of all-positive triples. This finding suggests that the processes of transitivity and balance are salient in all-positive triads, and this is consistent with Davis, Holland, and Leinhardt \cite{davis1979davis}'s observations for real-world small groups. We also found the occurrences of the one-negative triples to be higher than the all-negative triples, which is in contrast to several prior studies (\cite{szell2010multirelational,leskovec2010signed}). Further investigations are needed on the presence of the one-negative triad, as the likelihood of resolving the tension is likely to be unique across different relational types. 

Our findings are subject to several limitations. First of all, we only analyzed network structure and partial balance for the giant component of the networks in our sample. Our rationale is that only closed triads are considered for balance and transitivity analysis. Secondly, we consider partial balance only as a local property, at the triad level, of the network that has a global impact on the network. In future work, we will extend the measurement of partial balance and transitivity for longer cycles and examine if there are areas of tension detected in the network at other levels of analysis beyond the triad. 



\section*{Additional information}
The authors declare no competing interests.

\appendix

\section{Analysis of Emails using NLP} \label{appendix_1}
For the two email datasets, we used NLP methods to extract two types of edge signs from text data exchanged between the nodes (authors) that form a dyad: moral values (virtue or vice) and sentiment (positive or negative) \cite{rezapour2021user,dinh2023enhancing}. 
This approach is based on the premise that people's language use can reflect their cultural, economic, and ideological backgrounds \cite{triandis1989self}. 

To capture moral values in our email data sets, we leveraged the Moral Foundations Theory (MFT) \cite{graham2013moral, graham2009liberals}. MFT can help capture people's spontaneous reactions and categorize human behavior into five basic principles (fairness/cheating, care/harm, authority/subversion, loyalty/betrayal, and purity/degradation) that are characterized by opposing values (virtues and vices). 
To extract moral values from our email data, we used an enhanced version of MFD\footnote{\url{https://doi.org/10.13012/B2IDB-3805242_V1.1}} as developed, introduced, and validated in \cite{rezapour2019enhancing,atillinoisdatabankIDB-3957440}. 
The second NLP method we used for labeling links with signs is sentiment analysis; a technique commonly used for understanding people's opinions and affective state \cite{pang2008opinion}. 
To identify the sentiment of each email, we leverage a domain-adapted version of Subjectivity Lexicon, a widely used sentiment lexicon by \citet{wiebe2005creating}.

\bibliography{sample}

\end{document}